\documentclass{osa-article}

\journal{oe}

\usepackage{enumitem} 
\usepackage{booktabs}
\usepackage{threeparttable}
\usepackage{subfigure} 

\begin{document}
%
\title{Mathematical model and topology evaluation \\ of quantum key distribution network}

\author{Qiong~Li, Yaxing~Wang,\authormark{*} Haokun~Mao, Jiameng~Yao, and Qi~Han}

\address{Information Countermeasure Technique Institute, School of Computer Science and Technology, Harbin Institute of Technology, Harbin, 150080, China.}

\email{\authormark{*}yaxwang@163.com}



\begin{abstract}
Due to the intrinsic point-to-point characteristic of quantum key distribution (QKD) systems, it is necessary to study and develop QKD network technology to provide a secure communication service for a large-scale of nodes over a large area. Considering the quality assurance required for such a network and the cost limitations, building an effective mathematical model of a QKD network becomes a critical task. In this paper, a flow-based mathematical model is proposed to describe a QKD network using mathematical concepts and language. In addition, an investigation on QKD network topology evaluation was conducted using a unique and novel QKD network performance indicator, the \textsl{Information-Theoretic Secure communication bound}, and the corresponding linear programming-based calculation algorithm. A large number of simulation results based on the topologies of SECOQC network and NSFNET network validate the effectiveness of the proposed model and indicator.
\end{abstract}

\section{Introduction}
With the rapid development and increasing applicability of quantum key distribution (QKD) technology \cite{yin2016measurement, liao2017satellite, yuan201810, zhang2019continuous}, its intrinsic point-to-point feature \cite{bennett2014quantum} has become one of the major bottlenecks limiting the scale of its application. To overcome the limitation on the quantity of nodes \cite{tuysuz2011local} and communication distance, the construction of a QKD network with multiple QKD systems was an inevitable development trend. The QKD network in this paper is defined as a network that provides a secure communication service, utilizing the keys generated by QKD systems\cite{wang2019modeling}. In order to explore the physical feasibility of QKD networking, many practical QKD networks\cite{elliott2005current, elliott2007darpa, poppe2008outline, alleaume2009topological, fujiwara2011field, liao2018satellite} have been constructed in recent years. In the last decade, the number of nodes in existing QKD networks has expanded from 6\cite{poppe2008outline, sasaki2011field} to 56  nodes\cite{razavi2018introduction} and the communication distance has extended from 19.6\cite{peev2009secoqc} to 2000 km \cite{razavi2018introduction}. With the growing coverage and complexity of QKD networks, effective modeling is crucial for functional verification, quality assurance, cost control, cycle shortening, etc. \cite{dianati2008architecture,diamanti2016practical}.

The two primary approaches of network modeling include simulation models and mathematical models. Unlike traditional communication networks, the relevant research of QKD networks has not drawn much attention\cite{maurhart2013new, han2014novel,mahmud2019scaling, yang2017qkd,mehic2017implementation,wang2019modeling}. In 2017, Mehic et. al. designed a QKD network simulation model, QKDNetSim \cite{mehic2017implementation}, based on the classical Network Simulator-version 3 \cite{henderson2008network}, to evaluate and validate a network solution at a low cost. Although QKDNetSim could simulate the key generation and secure communication processes, it could not accurately reflect the practical performance of a QKD network, owing to its neglect of the actual key generation capability and volatile communication demand of the QKD network. In order to reflect the state of a practical network, we designed a practical QKD network simulation model in our previous work \cite{wang2019modeling}. In this model, the point-to-point key generation capability was modeled by the Gottesman-Lo-Lutkenhaus-Preskill (GLLP) theory \cite{gottesman2004security} and the volatile end-to-end communication demand was modeled by the Poisson stochastic process. Although our previous work enhanced the accuracy of the QKDNetSim, the inherent shortcomings of a simulation approach still exist, such as the empirical results, the difficult global optimal solution, etc.

A mathematical model, however, is a general mathematical abstraction of a QKD network, and therefore makes it possible to theoretically evaluate the performance of a QKD network and obtain the global optimal solution, etc. In the past two years, the problems of architecture \cite{tysowski2018engineering}, SDN \cite{aguado2019engineering, wang2019quantum}, routing \cite{huang2020quantum}, key management \cite{zhou2019security, cao2019kaas} and key allocation \cite{cao2019multi}, etc. are addressed. However, the results of our simulation model \cite{wang2019modeling} demonstrate that the performance of a practical QKD network primarily depends on how its key generation capability satisfies the communication demand. With emphasis on this characteristic, we are motivated to study the mathematical model of a QKD network, and its applications. 

\begin{itemize}[leftmargin=1em]
\item In this paper, a flow-based mathematical (FM) model is proposed. In the model, a QKD network was abstracted as the graph $G = \left( {V,E,F} \right)$, with the node set $V$, the edge set $E$ and the \textsl{QKD-flow} set $F$. According to the analysis of QKD network characteristics, the detailed attributes of the node and edge were analyzed. Furthermore, the \textsl{QKD-flow} was defined in reference to the generic \textsl{traffic-flow} \cite{schrijver2002history,han2014maximum}, which is a unique component of a QKD network, compared to a classical network.

\item Based on the FM model, an investigation on QKD network topology evaluation was conducted by proposing the indicator, the \textsl{Information-Theoretic Secure (ITS) communication bound}, and the corresponding linear programming-based calculation algorithm. The indicator is defined as the optimization of all whole demand satisfactions, to theoretically quantify the optimal performance of a QKD network topology \cite{calvert1997modeling}. The calculation of the \textsl{ITS communication bound} was inspired by the linear programming algorithm.

\item In order to verify the validity and necessity of the proposed FM model and performance indicator, two typical topology planning tasks, based on the existing topologies of SECOQC network\cite{peev2009secoqc} and NSFNET network\cite{yi2016provisioning}, were designed and analyzed. The simulation results demonstrated the advantages of the FM model and \textsl{ITS communication bound}.
\end{itemize}

The paper is organized as follows: In Section \ref{sec:related}, some related literature is discussed. In Section \ref{sec3}, the FM model is presented in detail. Based on the model, a unique QKD network performance indicator and the corresponding linear programming-based calculation algorithm are proposed in Section \ref{sec4}. In Section \ref{sec5}, the simulations of topology evaluation, based on the FM model, are presented and the results are analyzed. Section \ref{sec:conclusion} presents the concluding remarks.

\section{Related literature}
\label{sec:related}
In this section, related literature is reviewed and analyzed. The construction modes, application modes, and architecture models of a QKD network are discussed and the generic maximum-flow problem, which is usually used for task allocation, is introduced as one of the theoretical bases of the FM model.

\subsection{Construction modes of QKD network}
Construction modes used in existing QKD networks are divided into three main categories: optical switching, quantum relay, and trusted relay\cite{townsend1997quantum, kumavor2005comparison, ma2006polarization}. Because an optical switching device cannot break the scale limitation\cite{toliver2003experimental} and the core technique of quantum relay is still far from mature \cite{chen2017experimental, hu2019experimental}, the trusted relay is the most common construction mode at present.

\subsection{Application modes of QKD network}
Application modes used in existing QKD networks primarily include the key-by-key (also called key relay)\cite{yang2017qkd} and data-by-data (also called hop-by-hop)\cite{mehic2017implementation} modes. The main difference between these modes is how the communication is established. The key-by-key mode, used in the Tokyo network \cite{sasaki2011field}, can better retain the classical network protocol. However, in this mode, the number of key pools configured for each communication node is proportional to the number of potential communication parties. This requires a large memory capacity, which is impractical in a large-scale QKD network. In the data-by-data mode, which was used in the SECOQC network \cite{dianati2007architecture}, the number of key pools for each node is only related to the degree of the node \cite{bettstetter2002minimum}. This mode can greatly reduce the memory demand and, thus, increase the availability of a large-scale QKD network. Therefore, the data-by-data mode is more appropriate at present.

\subsection{Architecture models of QKD network}
\begin{figure}[htbp]
	\centering
	\includegraphics[scale=0.7]{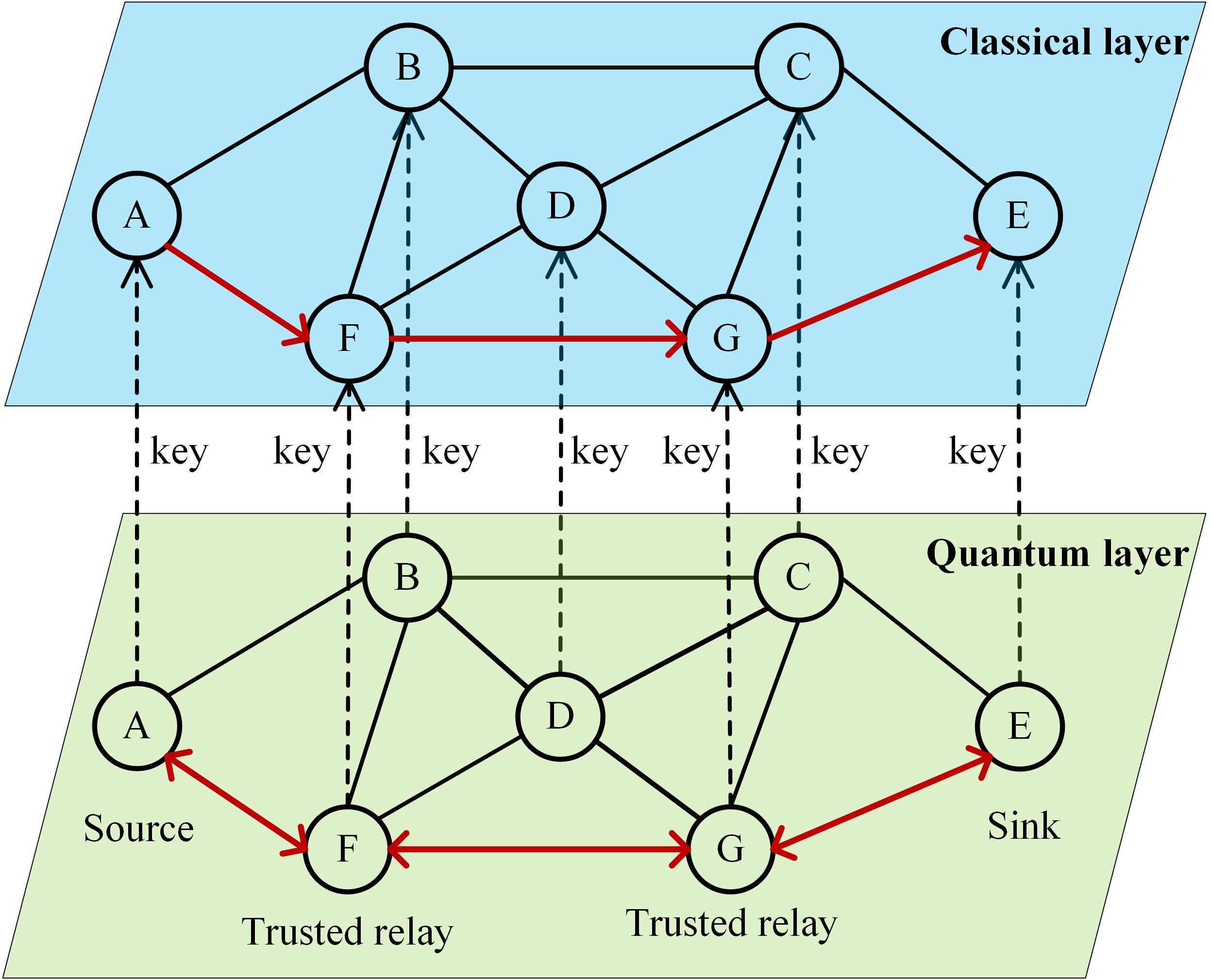}
	\caption{Two-layer architecture model of a QKD network\cite{wang2019modeling}}
	\label{fig_1} 
\end{figure}

Compared to a traditional communication network, the secure communication process between the end-to-end communication parties of a QKD network needs to consume the quantum keys generated by the point-to-point links. Therefore, a two-layer architecture model of a QKD network was proposed in our previous work \cite{wang2019modeling}, which is shown in Fig. \ref{fig_1}. In this model, end-to-end secure communication and point-to-point key generation proceed in the classical layer and the quantum layer, respectively.

Because the point-to-point key generation capability of a QKD system is extremely limited by the length of the quantum channel\cite{li2019improved, mao2019high} and is markedly lower than the capacity of the classical channel\cite{li2019high}, the performance of a QKD network is determined by the alignment of the communication demand and the key generation capability. Referring to our previous work\cite{wang2019modeling}, the point-to-point key generation capability of the quantum layer can be obtained by the common calculation method used for the secure key rate of a QKD system, such as GLLP theory \cite{gottesman2004security, li2016study} and the universal composable framework \cite{leverrier2015composable}. Assuming that the double decoy state protocol is adopted in the QKD network and the Chernoff bound\cite{curty2014finite} is used to estimate the finite code length effect\cite{ma2005practical, ma2012statistical}, the key generation capability ${R_{key}}$ can be calculated as,
\begin{equation}
\label{eq_R_k}
{R_{key}} = \max \left\{ {{f_{req}}{R_L},0} \right\},
\end{equation}
where ${R_L}$ represents the lower bound of the key generation capability for a photon, calculated as,
\begin{equation}
\label{eq_R_L}
{R_L} = - q{Q_\mu }{f_{ec}}H\left( {{E_\mu }} \right) + qQ_{_1}^L\left[ {1 - H\left( {e_{_1}^U} \right)} \right].
\end{equation}

In Eq. (\ref{eq_R_L}), $q$ is the sifting coefficient, the subscript $\mu $ denotes the intensity of the signal state, ${Q_\mu }$ is the overall gain of a signal state, ${E_\mu }$ is the overall quantum bit error rate, ${Q_{_1}^L}$ is the lower bound of the gain of single-photon state, ${e_{_1}^U}$ is the upper bound of the error rate of a single-photon state, ${f_{ec}}$ is the error correction efficiency, and $H\left( x \right)$ is the binary Shannon information function, given by $H\left( x \right) =  - x{\log _2}\left( x \right) - \left( {1 - x} \right){\log _2}\left( {1 - x} \right).$
To calculate ${R_L}$, four key variables, ${Q_\mu }$, ${E_\mu }$, ${Q_{_1}^L}$, and ${e_{_1}^U}$, are required. The first two can be directly measured through experiments and the latter two can be estimated by the decoy state method \cite{wang2019modeling}.

\subsection{The generic maximum-flow problem}
Let $G = \left( {V,E} \right)$ be a directed graph with node set $V$ and edge set $E$. The graph $G = \left( {V,E} \right)$ is a flow network\cite{goldberg1989network} if it has two distinguished nodes, a source $s \in V$, a sink $t \in V$, and a positive real-valued capacity $c\left( {u,v} \right)$ for each edge $\left( {u,v} \right) \in E$.\\

\textsl{Definition 1: A \textsl{traffic-flow} $f$ on $G$ is a non-negative function, ranging over all edges $\left( {u,v} \right) \in E$, satisfying the following constraints\cite{cormen2009introduction}:}
\begin{enumerate}[label=\textsl{(\roman *)}]
\item \textsl{Capacity constraint}
\begin{equation}
\label{traffic_capacity_capacity}
f\left( {u,v} \right) \le c\left( {u,v} \right), \forall \left( {u,v} \right) \in E.
\end{equation}

\item \textsl{Flow conservation}
\begin{equation}
\label{traffic_flow_conservation}
\sum\limits_{v \in V} {f\left( {u,v} \right)} - \sum\limits_{v \in V} {f\left( {v,u} \right)} = 0, \forall u \in V - \left\{ {s,t} \right\}.
\end{equation}
\end{enumerate}

Total value of a \textsl{traffic-flow}, $\left[\kern-0.15em\left[ f \right]\kern-0.15em\right]$, is defined as the total difference between the flows into and out of the sink $t$ \cite{goldberg1988new}, i.e.,
\begin{equation}
\label{eq_f}
\left[\kern-0.15em\left[ f \right]\kern-0.15em\right] = \sum\limits_{u \in V} {\left[ {f\left( {u,t} \right) - f\left( {t,u} \right)} \right]}.
\end{equation}
The maximum-flow problem aims to compute the maximum value of $\left[\kern-0.15em\left[ f \right]\kern-0.15em\right]$ for a given network and it is commonly discussed in the fields of the task assignment, logistics networks, urban planning, etc.

In the context of a communication network, there are usually multiple concurrent communication pairs, in the form of calls or connections\cite{brownlee1999rfc}. Therefore, the performance evaluation of a communication network is significantly more complicated than solving the maximum-flow problem. The classical solving algorithms for the maximum-flow problem, such as Ford-Fulkerson\cite{ford2009maximal} and Edmonds-Karp\cite{edmonds1972theoretical}, cannot be directly applied.

\section{Flow-based mathematical model}
\label{sec3}
With the increase in the coverage and complexity of existing QKD networks, it is beneficial to design an effective model for functional verification, quality assurance, cost control, cycle shortening, etc. In this section, a FM model is proposed. The ``flow'' does not refer to the generic \textsl{traffic-flow}, but the \textsl{QKD-flow}, which will be defined in \ref{sec_flow_conditions}.

By abstracting the communication party and trusted relay as nodes, the communication link as the edge, and the traffic volume as the \textsl{QKD-flow}, the definition of a QKD network is given below.\\

\textsl{Definition 2: A QKD network is modeled as a graph $G = \left( {V,E,F} \right)$, where $V $, $E$ and $F$ are the sets of nodes, edges, and \textsl{QKD-flows}, respectively.}

\subsection{Node attributes}
As a communication network, the most important task of a QKD network is to satisfy the communication demand between node pairs. The concept of \textsl{connection} is used to mathematically describe the communication demand.\\

\textsl{Definition 3: In the QKD network $G = \left( {V,E,F} \right)$, a \textsl{connection} $k_{ij} = \left( {{s_i},{t_j}} \right)$ $\left( {{s_i} \in V,{t_j} \in V} \right)$ indicates the communication demand between the node pair $\left( {{s_i},{t_j}} \right)$ \cite{cai2006link}, where ${s_i}$ is a source and ${t_j}$ is a sink.}\\

Let $K = \left\{ {\left( {{s_i},{t_j}} \right)|{s_i} \in V,{t_j} \in V} \right\}$ denote all the desired \textsl{connections} in the QKD network. Generally, the number of keys consumed in the communication process is determined by the communication demand and the key consumption ratio. The node attributes are illustrated in Table \ref{tab1}.
\begin{table}[htbp]
	\centering
	\caption{Attributes of node ${s_i}$}
	\label{tab1}      
	\begin{threeparttable}
    \renewcommand\arraystretch{1.5}
    \setlength{\tabcolsep}{1.5mm}{
        \begin{tabular}{ccc}
    		\toprule[1pt]
    		Attributes & Symbol & Value\\
    		\midrule
    		Communication demand & $d\left( {{s_i},{t_j}} \right)$ & $\left[ {0, + \infty } \right)$\\
            Key consumption ratio & $\beta \left( {{s_i},{t_j}} \right)$ & $\left[ {0,1} \right]$\\
    		\bottomrule[1pt]
    	\end{tabular}
    }
    \end{threeparttable}
\end{table}

The communication demand $d\left( {{s_i},{t_j}} \right)$ is the average communication rate required by the \textsl{connection} $\left( {{s_i},{t_j}} \right)$. Moreover, the communication demand of the node ${s_i}$ is denoted by $d\left( {s_i} \right) = \left\{ {d\left( {{s_i},{t_j}} \right)|{t_j} \in V} \right\}$.

The key consumption ratio $\beta \left( {{s_i},{t_j}} \right)$ is the ratio of the key length to the plaintext length in the adopted encryption algorithm. In particular, when the value of $\beta \left( {{s_i},{t_j}} \right)$ is 1, it indicates that a One-Time-Pad (OTP) algorithm\cite{watanabe2006security} was adopted to achieve secure communication. When the value of $\beta \left( {{s_i},{t_j}} \right)$ is 0, it indicates that the adopted encryption algorithm does not require the keys generated by the QKD systems. The key consumption ratio of the node ${s_i}$ is therefore denoted by $\beta \left( {s_i} \right) = \left\{ {\beta \left( {{s_i},{t_j}} \right)|{t_j} \in V} \right\}$.

\subsection{Edge attributes}
Because upstream and downstream communication share channel bandwidth \cite{sklar1988digital}, the edge of a QKD network is considered undirected. The undirected edge, formed by connecting nodes ${u_m} \in V$ and ${v_n} \in V$, is denoted by $\left( {{u_m},{v_n}} \right) \in E$.

The main attribute of a QKD network lies in the fact that the key generation process requires the participation of a quantum channel and the key generation rate is very limited by the length of the quantum channel. In order to mathematically describe this characteristic, several important attributes of the edge are extracted. Their symbol representations and value ranges are shown in Table \ref{tab2}.
\begin{table}[htbp]
	\centering
	\caption{Attributes of the edge $\left( {{u_m},{v_n}} \right)$}
	\label{tab2}       
	\begin{threeparttable}
    \renewcommand\arraystretch{1.5}
    \setlength{\tabcolsep}{1.5mm}{
        \begin{tabular}{ccc}
    		\toprule[1pt]
    		Attributes & Symbol & Value\\
    		\midrule
            Classical channel capacity & $c\left( {{u_m},{v_n}} \right)$ & $\left[ {0, + \infty } \right)$\\
            Key generation capability & $r\left( {{u_m},{v_n}} \right)$ & $\left[ {0, + \infty } \right)$\\
    		\bottomrule[1pt]
    	\end{tabular}
    }
    \end{threeparttable}
\end{table}

Classical channel capacity $c\left( {{u_m},{v_n}} \right)$ represents the capability of a classical channel to transmit information. When $c\left( {{u_m},{v_n}} \right)$ is 0, there is no classical channel on the edge $\left( {{u_m},{v_n}} \right)$, resulting in the infeasibility of a secure communication process.

Key generation capability $r\left( {{u_m},{v_n}} \right)$ is related to
the parameters of the QKD system configured on the edge $\left( {{u_m},{v_n}} \right)$. In particular, when the $c\left( {{u_m},{v_n}} \right)$ is 0, there is no classical channel on the edge $\left( {{u_m},{v_n}} \right)$. Because the classical channel is required for the transmission of supplementary information during the key exchange \cite{li2015efficient}, the key generation process cannot proceed on this edge. Suppose decoy state discrete-variable QKD systems are configured, according to Eq. (\ref{eq_R_k}), the key generation capability of the edge $\left( {{u_m},{v_n}} \right)$ is calculated as,
\begin{equation}
\label{eq_r}
r\left( {{u_m},{v_n}} \right) = \left\{ \begin{array}{l}
R_{key},\quad \ \  c\left( {{u_m},{v_n}} \right) \ne 0\\
0, \quad \quad \quad   c\left( {{u_m},{v_n}} \right) = 0.
\end{array} \right.
\end{equation}

\subsection{Flow conditions}
\label{sec_flow_conditions}
Although the concept of \textsl{traffic-flow} is referred to in this paper, the flow in a QKD network has many unique features owing to the significant difference between a QKD network and generic flow network. For example, there exist many \textsl{connections} and the edge owns two types of capacities.\\

\textsl{Definition 4: In a QKD network $G = \left( {V,E,F} \right)$, a \textsl{QKD-flow} $f \in F$ is a non-negative function ranging over all \textsl{connections} $\left( {{s_i},{t_j}} \right) \in K$ and all edges $\left( {{u_m},{v_n}} \right) \in E$, which is represented by a symbol $f\left( {{s_i},{t_j},{u_m},{v_n}} \right)$.}\\

Specifically, a \textsl{QKD-flow} $f\left( {{s_i},{t_j},{u_m},{v_n}} \right)$ characterizes the net data flow from a sources $s_i$ to a sink $t_j$ on an edge $(u_m, v_n)$. In other words, two \textsl{QKD-flows} are different when their sources are different or their sinks are different or their edges are different.

The \textsl{QKD-flow} set $F$ can be written as $F = \left\{ {f\left( {{s_i},{t_j},{u_m},{v_n}} \right)|\left( {{s_i},{t_j}} \right) \in K,\left( {{u_m},{v_n}} \right) \in E} \right\}$. Because the secure transmission process is organized in packets, the value of $f\left( {{s_i},{t_j},{u_m},{v_n}} \right)$ must be in integer multiples of the packet size $P$, which is called numerical constraint and given as,

\begin{equation}
\label{eq_flow_value}
 \frac{{f\left( {{s_i},{t_j},{u_m},{v_n}} \right)}}{P} \in N, \forall \left( {{u_m},{v_n}} \right) \in E, \forall \left( {{s_i},{t_j}} \right) \in K,
\end{equation}
where $N$ is the set of all non-negative integers.

When the value of $f\left( {{s_i},{t_j},{u_m},{v_n}} \right)$ is 0, it indicates that there is no flow of the \textsl{connection} $\left( {{s_i},{t_j}} \right)$ on the edge $\left( {{u_m},{v_n}} \right)$. In addition, the secure communication process is directed, therefore, $f\left( {{s_i},{t_j},{u_m},{v_n}} \right)$ is considered a directed flow. Therefore, $f\left( {{s_i},{t_j},{u_m},{v_n}} \right)$ and $f\left( {{s_i},{t_j},{v_n},{u_m}} \right)$ are different. The special conditions of the \textsl{QKD-flow} set are analyzed below.\\

\begin{enumerate}[label=\textsl{(\roman *)}]
\item \textsl{Capacity constraint}
\begin{itemize}[leftmargin=0em]
\item[-]{
For all $\left( {{u_m},{v_n}} \right) \in E$, the total flow on the edge $\left( {{u_m},{v_n}} \right)$ and its reverse edge $\left( {{v_n},{u_m}} \right)$ must be non-negative and less than or equal to its classical channel capacity. In addition, as an undirected graph, the classical channel capacity is shared by upstream and downstream flows. Thus, Eq. (\ref{eq_capacity1}) should be satisfied.}

\begin{equation}
\label{eq_capacity1}
\begin{split}
0 & \le \sum\limits_{\left( {{s_i},{t_j}} \right) \in K} {\left[ {f\left( {{s_i},{t_j},{u_m},{v_n}} \right) + f\left( {{s_i},{t_j},{v_n},{u_m}} \right)} \right]}\\
  & \le c\left( {{u_m},{v_n}} \right),\forall ( {{u_m},{v_n}} ) \in E.
\end{split}
\end{equation}
\item[-]{
For all $\left( {{u_m},{v_n}} \right) \in E$, the total key consumption on the edge $\left( {{u_m},{v_n}} \right)$ and its reverse edge $\left( {{v_n},{u_m}} \right)$ must be non-negative and less than or equal to its key generation capability. Considering the key consumption ratio $\beta \left( {{s_i},{t_j}} \right)$, the relationship between the total flow on the edge $\left( {{u_m},{v_n}} \right)$ and its reverse edge $\left( {{v_n},{u_m}} \right)$ is given by,}
\begin{equation}
\label{eq_capacity2}
\begin{aligned}
0 & \le \sum\limits_{\left( {{s_i},{t_j}} \right) \in K} {\beta \left( {{s_i},{t_j}} \right)\left[ {f\left( {{s_i},{t_j},{u_m},{v_n}} \right)+f\left( {{s_i},{t_j},{v_n},{u_m}} \right)} \right]}\\
  & \le r\left( {{u_m},{v_n}} \right), \forall \left( {{u_m},{v_n}} \right) \in E,
\end{aligned}
\end{equation}

\end{itemize}

\item \textsl{Flow conservation}
\begin{itemize}[leftmargin=0em]
\item[-]{
    For all \textsl{connections} $\left( {{s_i},{t_j}} \right) \in K$ and all non-source and non-sink nodes ${u_m} \in V - \left\{ {{s_i},{t_j}} \right\}$, the total flow into the node ${u_m}$ must equal to the total flow out of it, i.e.,}
\begin{equation}
\label{eq_flow}
\begin{aligned}
\sum\limits_{{v_n} \in V} {f\left( {{s_i},{t_j},{u_m},{v_n}} \right)} - \sum\limits_{{v_n} \in V} {f\left( {{s_i},{t_j},{v_n},{u_m}} \right)} = 0, \\
\forall {u_m} \ne {s_i},{t_j},\forall \left( {{s_i},{t_j}} \right) \in K.
\end{aligned}
\end{equation}
\end{itemize}
\end{enumerate}

In general, a FM model suitable for a QKD network was proposed by adding the attributes of key consumption ratio and key generation capability and giving the definition and conditions of \textsl{QKD-flows}. As the theoretical foundation of topology evaluation and design, routing evaluation and design, QKD systems selection, and construction cost control, the FM model can be used not only for the construction of a new QKD network, but also for the optimization of existing QKD networks.

\section{FM model based topology evaluation}
\label{sec4}
To construct a high performance QKD network, designing a precise topology evaluation scheme is one of the most important tasks. An investigation on QKD network topology evaluation was conducted, based on the FM model. Firstly, the \textsl{ITS communication bound} indicator was designed to mathematically describe the quality of QKD network topology. In addition, a linear programming-based calculation algorithm is proposed to obtain the quantitative quality.

\subsection{The description of topology quality}
To eliminate the influence of the encryption algorithm on the topology evaluation, an OTP algorithm\cite{watanabe2006security}, which can provide an ITS communication service, is adopted in this section. Hence, for all $\left( {{s_i},{t_j}} \right) \in K$, the value of $\beta \left( {{s_i},{t_j}} \right)$ is set to 1. The quality of a QKD network topology is measured by the proposed performance indicator \textsl{ITS communication bound}.

For a given \textsl{connection} $\left( {{s_i},{t_j}} \right)$, similar to \textsl{traffic-flow}, its total value, $\left[\kern-0.15em\left[{f\left( {{s_i},{t_j}} \right)} \right]\kern-0.15em\right] $, is the total difference between the \textsl{QKD-flows}, which belong to the {connection} $\left( {{s_i},{t_j}} \right)$, into and out of the sink ${t_j}$, represented as,
\begin{equation}
\label{eq_fst}
\left[\kern-0.15em\left[{f\left( {{s_i},{t_j}} \right)} \right]\kern-0.15em\right] = \sum\limits_{{u_m} \in V,{v_n} = {t_j}} {\left[ {f\left( {{s_i},{t_j},{u_m},{v_n}} \right) - f\left( {{s_i},{t_j},{v_n},{u_m}} \right)} \right]}.
\end{equation}

Let $M\left( {{s_i},{t_j}} \right)$ represent the connection demand satisfaction for a given \textsl{connection} $\left( {{s_i},{t_j}} \right)$, which is the ratio of its total value to its communication demand, i.e.,
\begin{equation}
\label{eq_m}
M\left( {{s_i},{t_j}} \right) = \frac{{\left[\kern-0.15em\left[{f\left( {{s_i},{t_j}} \right)} \right]\kern-0.15em\right] }}{{d\left( {{s_i},{t_j}} \right)}}.
\end{equation}
$M\left( {{s_i},{t_j}} \right) \ge 1$ indicates that the communication demand of this connection $\left( {{s_i},{t_j}} \right)$ is satisfied. 

Then, let's calculate all connection demand satisfactions for all connections. Next, let's obtain the minimum value from all connection demand satisfactions. This minimum value can be used to represent a whole demand satisfaction of the given QKD network. Obviously, each possible \textsl{QKD-flow} set corresponds to a specific whole demand satisfaction. Similar to the generic maximum-flow problem, we can obtain the maximum value of all possible whole demand satisfactions through the optimization of \textsl{QKD-flows}. This maximum value, which is called \textsl{ITS communication bound}, can be used to represent the optimal performance of a QKD network with a given topology.
\\

\textsl{Definition 5: For a QKD network with a given topology, the \textsl{ITS communication bound} is defined as the optimal (maximal) whole demand satisfaction, which is defined as the minimum of all connection demand satisfactions for all \textsl{connections} in $K$ over a possible \textsl{QKD-flow}.}\\

The \textsl{ITS communication bound} $B$ is therefore can be calculated as,

\begin{subequations}
\begin{align}
\label{eq_B}
&B = \mathop {\max }\limits_F {\kern 1pt} \rho \left( F \right),\\
&\rho\left( F \right) =  {\min_{\left( {{s_i},{t_j}} \right) \in K}{ M\left( {{s_i},{t_j}} \right)}},
\end{align}
\end{subequations}
where $ \rho\left( F \right) $ is the minimum value of all connection demand satisfactions when the \textsl{QKD-flow} set is $F$.

It is clear that the communication demand for all \textsl{connections} are satisfied only when the value of $B$ is greater than 1. The larger the value of $B$ , the higher the degree of satisfaction. It is also significant that the gap between the performance of a specific QKD network with specific routing protocols and specific key management strategies, and the calculation of \textsl{ITS communication bound} of this QKD network can be used to evaluate the performance of the routing protocols and key management strategies.

\subsection{The calculation of topology quality}
To calculate the indicator $B$, it is necessary to explore the optimal assignment of the \textsl{QKD-flows}, which is defined as the multi-connection flow problem (MCFP) in this paper. Although the MCFP appears to be a combination of several maximum-flow problems, the interaction of multiple maximum-flow problems cause their respective solutions to fail \cite{dai2016finding}.

With regard to the definition of the \textsl{QKD-flow}, the flow must satisfy the capacity constraint and flow conservation. Therefore, the MCFP can be formulated as,
\begin{subequations}
\label{eq_original}
\begin{alignat}{2}
\max_{F} \quad & \min_{{\left( {{s_i},{t_j}} \right) \in K}} {\frac{{\sum\limits_{{u_m} \in V,{v_n} = {t_j}} {\left[ {f\left( {{s_i},{t_j},{u_m},{v_n}} \right) - f\left( {{s_i},{t_j},{v_n},{u_m}} \right)} \right]}}}{{d\left( {{s_i},{t_j}} \right)}}},\\
\mbox{s.t.} \quad
   &   Eq. \left(\ref{eq_flow_value}\right), Eq. \left(\ref{eq_capacity1}\right), Eq. \left(\ref{eq_capacity2}\right), Eq. \left(\ref{eq_flow}\right).
\end{alignat}
\end{subequations}

In Eq. (\ref{eq_original}), $F = \left\{ {f\left( {{s_i},{t_j},{u_m},{v_n}} \right)|\left( {{s_i},{t_j}} \right) \in K,\left( {{u_m},{v_n}} \right) \in E} \right\}$ are the set of decision variables. The formulation is very similar to that of the linear programming problem. However, due to the issues of a non-linear objective function and the non-standard data type of the decision variables, the MCFP is not a standard linear programming problem, which is difficult to solve. To transform this problem into standard linear programming, the original decision variable $f\left( {{s_i},{t_j},{u_m},{v_n}} \right)$ must be converted into a new variable $x\left( {{s_i},{t_j},{u_m},{v_n}} \right)$ as follows:
\begin{equation}
\label{eq_x}
x\left( {{s_i},{t_j},{u_m},{v_n}} \right) = \frac{{f\left( {{s_i},{t_j},{u_m},{v_n}} \right)}}{P}.
\end{equation}
Therefore, $X = \left\{ {x\left( {{s_i},{t_j},{u_m},{v_n}} \right)|\left( {{s_i},{t_j}} \right) \in K,\left( {{u_m},{v_n}} \right) \in E} \right\}$ becomes the new set of decision variables. In addition, the original objective function is replace by a new objective function $\rho\left( F \right)$, by adding $\rho\left( F \right)$ as a new decision variable and adding two constraint conditions, as follows,
\begin{subequations}
\label{eq_BP}
\begin{align}
& \frac{\rho\left( F \right)}{P} - \frac{{\sum\limits_{{u_m} \in V,{v_n} = {t_j}} {\left[ {x\left( {{s_i},{t_j},{u_m},{v_n}} \right) - x\left( {{s_i},{t_j},{v_n},{u_m}} \right)} \right]}}}{{d\left( {{s_i},{t_j}} \right)}} \le 0, \forall \left( {{s_i},{t_j}} \right) \in K,\\
& \rho\left( F \right) \in R_0^ + ,
\end{align}
\end{subequations}
where $R_0^ +$ is the set of non-negative real numbers.

According to the above operations, the MCFP is transformed into an equivalent standard mixed integer linear-programming problem, which is formulated as: 

\begin{subequations}
\label{eq_new}
\begin{align}
    \max_{X} \quad & \rho\left( F \right)\\
    \mbox{s.t.} \quad
    & 0 \le \sum\limits_{\left( {{s_i},{t_j}} \right) \in K} {x\left( {{s_i},{t_j},{u_m},{v_n}} \right)} + \sum\limits_{\left( {{s_i},{t_j}} \right) \in K} {x\left( {{s_i},{t_j},{v_n},{u_m}} \right)} \le \frac{{c\left( {{u_m},{v_n}} \right)}}{P}, \forall \left( {{u_m},{v_n}} \right) \in E,\\
    & 0 \le \sum\limits_{\left( {{s_i},{t_j}} \right) \in K} {x\left( {{s_i},{t_j},{u_m},{v_n}} \right)} + \sum\limits_{\left( {{s_i},{t_j}} \right) \in K} {x\left( {{s_i},{t_j},{v_n},{u_m}} \right)} \le \frac{{r\left( {{u_m},{v_n}} \right)}}{P}, \forall \left( {{u_m},{v_n}} \right) \in E,\\
    & \sum\limits_{{v_n} \in V} {x\left( {{s_i},{t_j},{u_m},{v_n}} \right)} - \sum\limits_{{v_n} \in V} {x\left( {{s_i},{t_j},{v_n},{u_m}} \right)} = 0, \forall {u_m} \ne {s_i}{\rm{,}}{t_j},\forall \left( {{s_i},{t_j}} \right) \in K,\\
    & \frac{\rho\left( F \right)}{P} - \frac{{\sum\limits_{{u_m} \in V,{v_n} = {t_j}} {\left[ {x\left( {{s_i},{t_j},{u_m},{v_n}} \right) - x\left( {{s_i},{t_j},{v_n},{u_m}} \right)} \right]} }}{{d\left( {{s_i},{t_j}} \right)}} \le 0, \forall \left( {{s_i},{t_j}} \right) \in K,\\
    & x\left( {{s_i},{t_j},{u_m},{v_n}} \right) \in N, \forall \left( {{u_m},{v_n}} \right) \in E,\forall \left( {{s_i},{t_j}} \right) \in K,\\
    & \rho\left( F \right) \in R_0^ + .
\end{align}
\end{subequations}

In order to solve this problem, a linear programming solver\cite{meindl2012analysis}, Gurobi\cite{optimization2014inc}, was adopted. In the formulation, $r\left( {{u_m},{v_n}} \right)$ represents the key generation capability of a QKD system. It is important to note that many types of QKD systems can be adopted into the QKD network and the corresponding topology quality can be obtained by changing the calculation of $r\left( {{u_m},{v_n}} \right)$.

\section{Simulation results and analysis}
\label{sec5}
\subsection{Simulation design}
The design of this simulation consisted mainly of the design of the parameters for communication demand, QKD systems, packet size, and classical channel capacities. These parameters will directly affect the topology performance of a specific QKD network. During the simulation, typical topologies of a SECOQC and NSFNET network were adopted.

To simplify the analysis, the communication demand between any two different nodes in the simulation was assumed to be the same, denoted as $d$, i.e.,
\begin{subequations}
\label{eq_K}
\begin{align}
&K = \left\{ {\left( {{s_i},{t_j}} \right)|{s_i} \in V,{t_j} \in V,{s_i} \ne {t_j}} \right\},\\
&d\left( {{s_i},{t_j}} \right) = \left\{ \begin{array}{l}
d,\left( {{s_i},{t_j}} \right) \in K,\\
0,\left( {{s_i},{t_j}} \right) \notin K.
\end{array} \right.
\end{align}
\end{subequations}

Given that the discrete-variable QKD protocol is one of the most practical QKD protocols and that the decoy state method is critical to security assurance, a decoy state discrete-variable QKD system was adopted in this simulation. To simplify analysis, the parameters of all QKD systems were assumed to be the same. To facilitate a comparison with the performance reported in the literature\cite{wang2019modeling}, the same parameters of QKD system, listed in Table \ref{tab3}, and the same packet size, that is, $P = 500$ bytes, were adopted in this simulation. 

In Table \ref{tab3}, ${f_{req}}$ is the repetition rate, $q$ is the sifting coefficient, $\alpha$ is the fiber attenuation coefficient, ${\eta _{Bob}}$ is the transmittance of Bob, $e_{det}$ is the intrinsic error rate due to misalignment and instability of the optical system, $\mu$ is the intensity of signal state, $\nu$ is the intensity of decoy state, $\phi$ is the intensity of vacuum state, ${Y_0}$ is the background rate, ${e_0}$ is the error rate of the background, ${f_{ec}}$ is the error correction efficiency, ${N_\mu}$ is the number of signal pluses sent by Alice, ${N_\nu}$ is the number of decoy pluses sent by Alice, ${N_\phi}$ is the number of vacuum pluses sent by Alice, $\varsigma$ is the security bound.

\begin{table*}[htbp]
	\centering
	\caption{Parameters of QKD systems\cite{wang2019modeling}}
	\label{tab3}       
	\begin{threeparttable}
    \renewcommand\arraystretch{1.5}
    \setlength{\tabcolsep}{1.2mm}{
        \begin{tabular}{ccccccccccccccc}
    		\toprule[1pt]
    		${f_{req}}$ & $q$ & $\alpha $ & $\eta_{Bob}$ & $e_{det}$ & $\mu$ & $\nu$ & $\phi$ & $Y_0$ & $e_0$ & ${f_{ec}}$ & ${N_{\mu}}$ & ${N_{\nu}}$ & ${N_{\phi}}$ & $\varsigma$ \\
    		\midrule
    		1GHz & 0.9 & 0.2dB/km & 0.1 & 0.01 & 0.4 & 0.1 & 0 & 2.1E-5 & 0.5 & 1.15 & 1.6E10 & 2E9 & 2E9 & 5.73E-7\\
    		\bottomrule[1pt]
    	\end{tabular}
    }
    \end{threeparttable}
\end{table*}

Because classical optical fiber communication technology is sufficiently mature \cite{hecht2011ultrafast, idler2017field
}, the classical channel capacities of all the edges were set to 1 Gbps, i.e.,
\begin{equation}
\label{eq_c}
c\left( {{u_m},{v_n}} \right) = 1{\ \rm{Gbps}}, \forall \left( {{u_m},{v_n}} \right) \in E.
\end{equation}

\subsection{Topology evaluation based on QKD systems placement}
A typical task undertaken as part of network topology planning involves investigating how to effectively enhance the network performance by adding just one system to existing topology. In the context of a QKD network, the equivalent task involves finding out how to effectively enhance the QKD network performance by adding a QKD system to an existing QKD network topology. When the new QKD system is added, which is equivalent to adding a new edge to existing topology, a modified topology is actually formed.

It is well known that the key generation process is reliant on the optical fiber. Therefore, a QKD system can function only when placed on the existing edge. In addition, adding a QKD system to the edge $\left( {{u_m},{v_n}} \right)$ results in an increase in the key generation capability of this edge and its reverse edge, $r\left( {{u_m},{v_n}} \right)$ and $r\left( {{v_n},{u_m}} \right)$.
Due to the different link distances, the key generation capabilities of a given QKD system will vary depending on the edge on which it is placed. In addition, due to the different traffic burdens of edges within the overall network, any given QKD system will produce different gains in communication security depending on the edge on which it is placed.

\begin{figure}[htbp]
	\centering
	\includegraphics[scale=0.5]{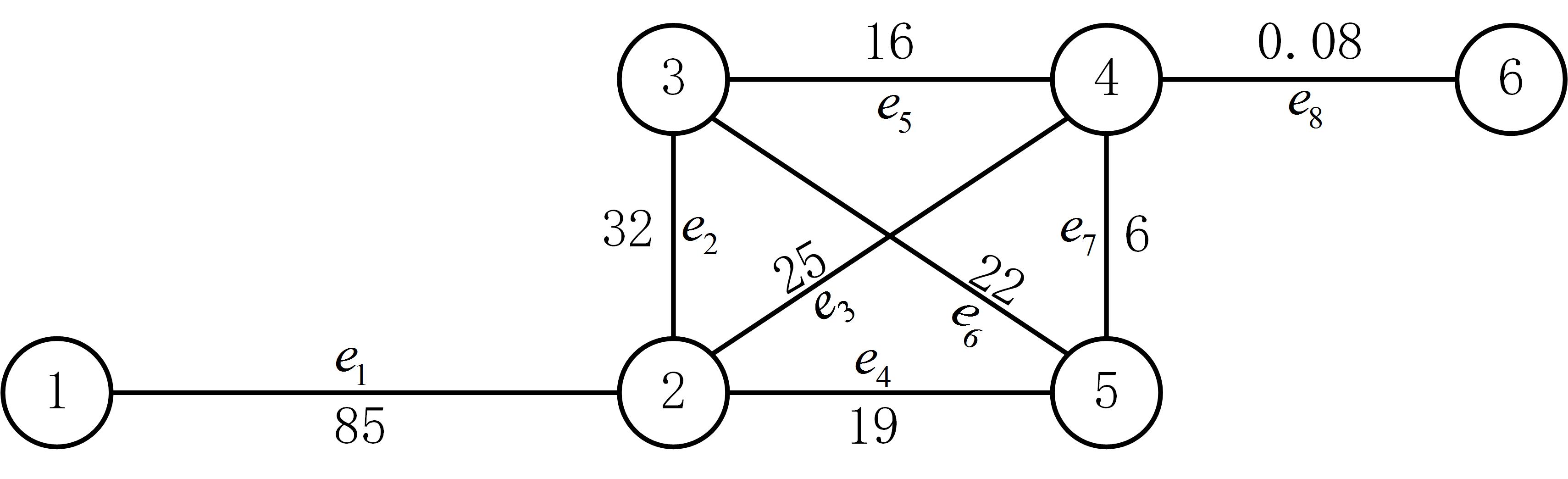}
	\caption{Topology of SECOQC network\cite{wang2019modeling}}
	\label{fig_2} 
\end{figure}

To verify validity of the proposed FM model and performance indicator, a network topology of the SECOQC network, which is shown in Fig. \ref{fig_2}, was selected. The numbers on the edges in the figure represent the link distances in kilometers. 

To determine the optimal placement scheme, the QKD system is placed at every possible edge in a SECOQC network, to form eight modified topologies. To quantitatively evaluate different placement schemes, the \textsl{ITS communication bounds} of the original SECOQC topology and the eight modified SECOQC topologies were calculated. The results are listed in Table \ref{tab4}.

Table \ref{tab4} shows that the \textsl{ITS communication bound} increased and the communication demand switched from unsatisfied to satisfied when the QKD system was placed on edge ${e_1}$. This indicates that edge ${e_1}$ acts as a bottleneck in the topology of the SECOQC network, which is consistent with the performance results in the literature\cite{wang2019modeling}.
\begin{table}[htbp]
	\centering
	\caption{Performance comparison of original SECOQC topology and modified SECOQC topologies $\left( {d = 25{\ \rm{Kbps}}} \right)$}
	\label{tab4}       
	\begin{threeparttable}
    \renewcommand\arraystretch{1.5}
    \setlength{\tabcolsep}{1.5mm}{
        \begin{tabular}{cccccccccc}
    		\toprule[1pt]
    		Placement & $none$ & ${e_1}$ & ${e_2}$ & ${e_3}$ & ${e_4}$ & ${e_5}$ & ${e_6}$ & ${e_7}$ & ${e_8}$\\
    		\midrule
    		Bound & 0.96 & 1.92 & 0.96 & 0.96 & 0.96 & 0.96 & 0.96 & 0.96 & 0.96\\
    		\bottomrule[1pt]
    	\end{tabular}
    }
    \end{threeparttable}
\end{table}

If we observe the original SECOQC topology in Fig. \ref{fig_2}, we can see that the length of edge ${e_1}$ is 85 km. By substituting this length into Eq. (\ref{eq_R_L}), the calculated key generation capability is about 233 Kbps. However, as a ``bridge'' \cite{mchugh1990algorithmic} in the topology, no matter which routing algorithm is adopted, the keys generated on edge ${e_1}$ must satisfy the communication demand for ten connections $\left( {{s_1},{t_2}} \right)$, $\left( {{s_1},{t_3}} \right)$, $\left( {{s_1},{t_4}} \right)$, $\left( {{s_1},{t_5}} \right)$, $\left( {{s_1},{t_6}} \right)$, $\left( {{s_2},{t_1}} \right)$, $\left( {{s_3},{t_1}} \right)$, $\left( {{s_4},{t_1}} \right)$, $\left( {{s_5},{t_1}} \right)$ and $\left( {{s_6},{t_1}} \right)$ . Thus, when the communication demand is set to 25 Kbps, the \textsl{ITS communication bound} can be calculated as ${\text{233/(25*10)}} \approx {\text{0.93}}$. In addition, when the addition QKD system is placed on the edge ${e_1}$, the modified \textsl{ITS communication bound} can be calculated as ${\text{(233*2)/(25*10)}} \approx {\text{1.86}}$. The alignment between this calculation and the result shown in Table \ref{tab4} is due to the numerical constraint of \textsl{QKD-flow}.

It is clear that the distances corresponding to the other edges are significantly shorter than edge ${e_1}$, which is conducive to a higher key generation capability. As the traffic burden is eased, only one QKD system placed on these edges will not exhibit a more significant performance improvement. The validity of the FM model and the \textsl{ITS communication bound} indicator is verified.

\subsection{Topology evaluation based on intermediate nodes selection}

Given that above results are too obvious to prove the validity of the research, another simulation based on intermediate nodes selection was designed to further verify the necessity of the proposed model and indicator.

\begin{figure}[htbp]
	\centering
	\includegraphics[scale=0.5]{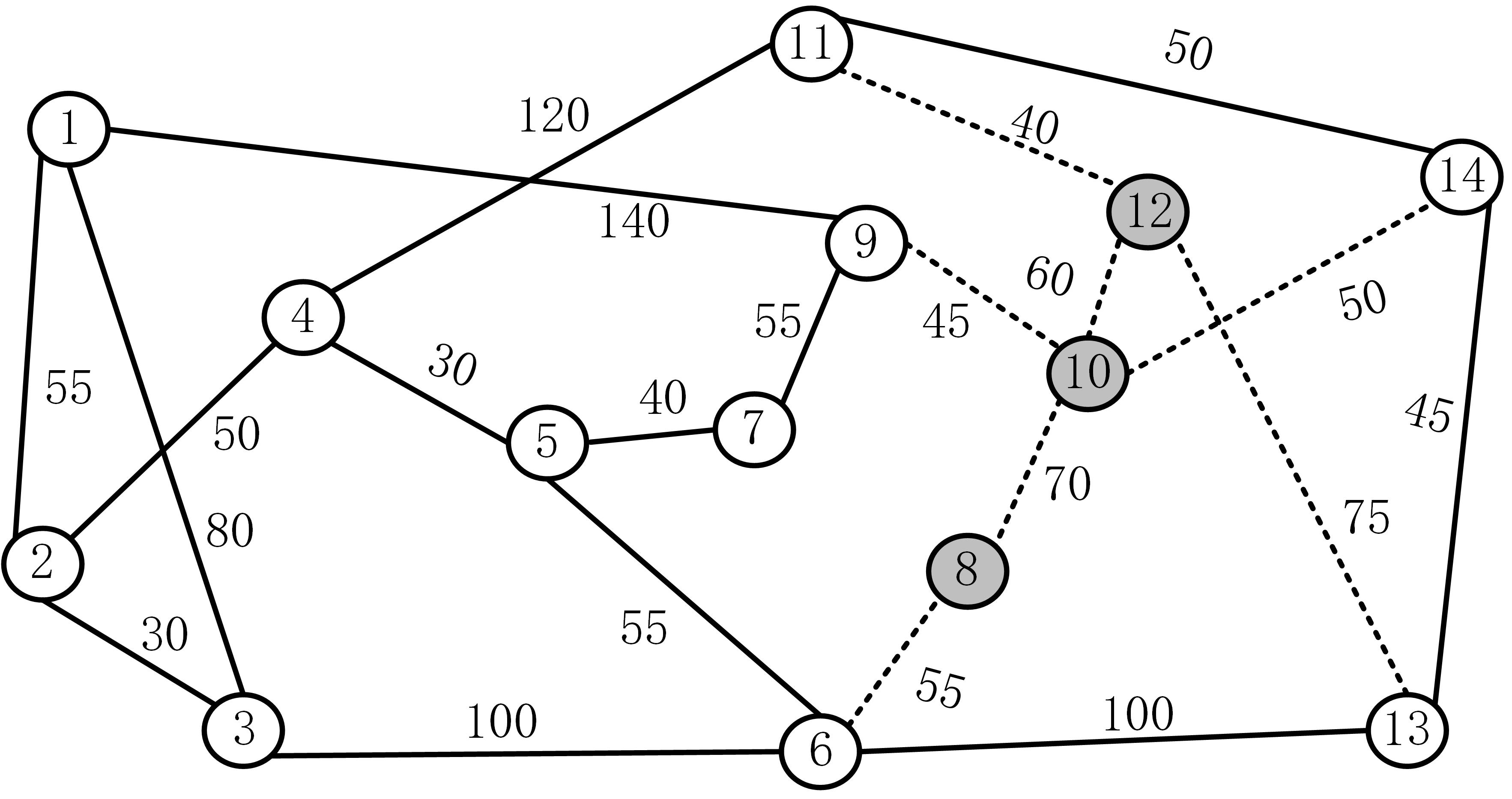}
	\caption{Topology of NSFNET network}
	\label{fig_3} 
\end{figure}

Nodes selection, investigating how to effectively enhance the network performance by adding several intermediate nodes and corresponding edges to existing topology, is also a typical task in network topology planning. Since a QKD network based on trusted relay mode requires every node in the topology to be trusted, the addition of each new node will reduce the security of this QKD network. Therefore,  compared with a classical network, nodes selection is more important for a QKD network. To verify necessity of the proposed FM model and performance indicator, a complicated topology of the NSFNET network, which is shown in Fig. \ref{fig_3}, was designed. 

Three gray nodes in Fig. \ref{fig_3} are optional nodes with no communication demand. In contrast, every pair of other nodes have the same communication demand. A dotted line in the figure represents an optional edge that connects one or two optional nodes. To simplify analysis, this simulation assumed that an optional edge is selected if and only if the nodes at both ends of this edge are selected. In addition, the same parameters for communication demand, QKD system, packet size, and classical channel capacities were adopted.

\begin{figure}[htbp]
\centering
	\subfigure[Original NSFNET topology]{
	\label{fig:nsfnet1}
	\includegraphics[scale=0.4]{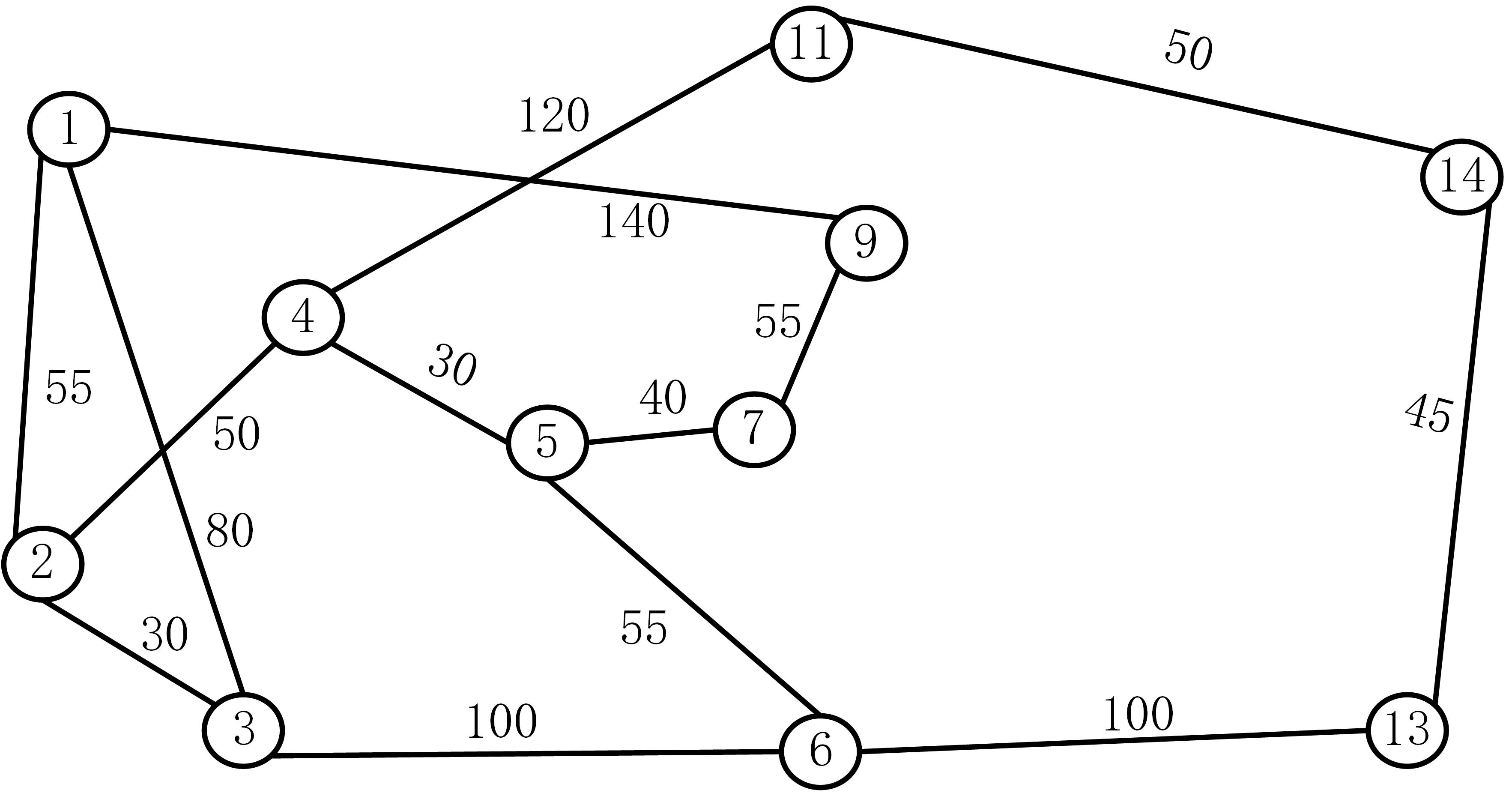}}
	\subfigure[Adding node ${v_8}$]{
	\label{fig:nsfnet2}
	\includegraphics[scale=0.4]{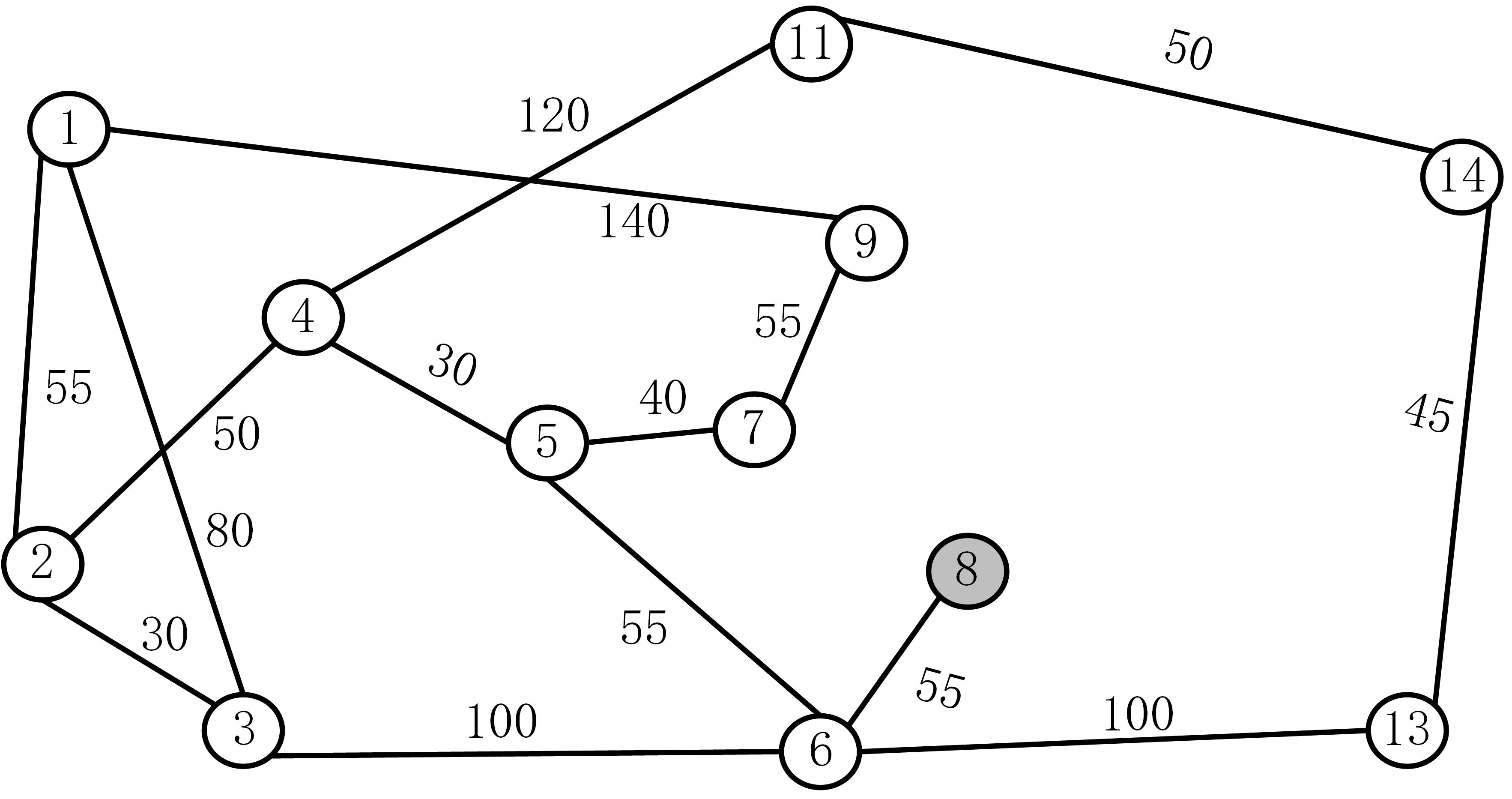}}

	\subfigure[Adding node ${v_{10}}$]{
	\label{fig:nsfnet3}
	\includegraphics[scale=0.4]{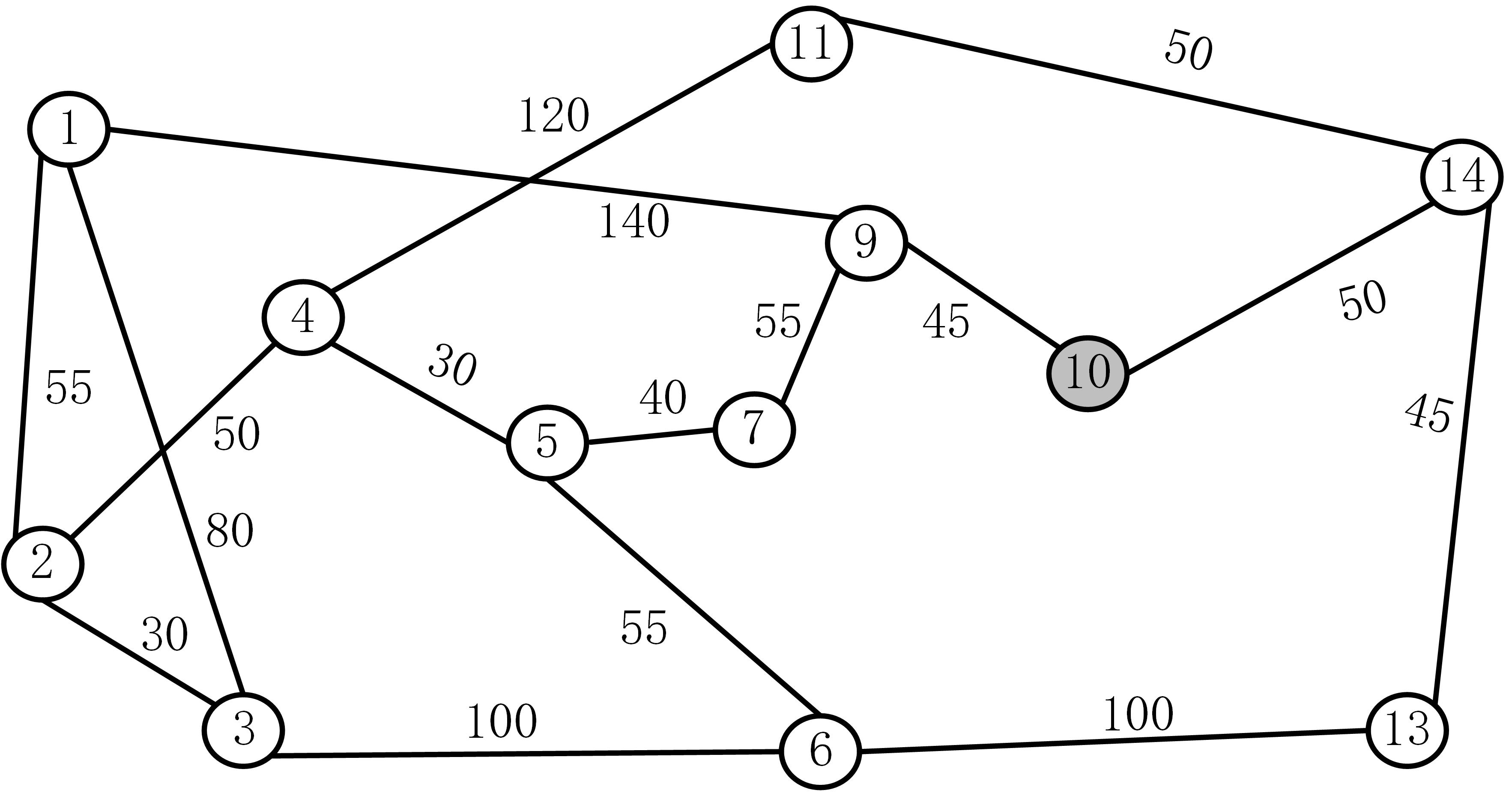}}
	\subfigure[Adding node ${v_{12}}$]{
	\label{fig:nsfnet4}
	\includegraphics[scale=0.4]{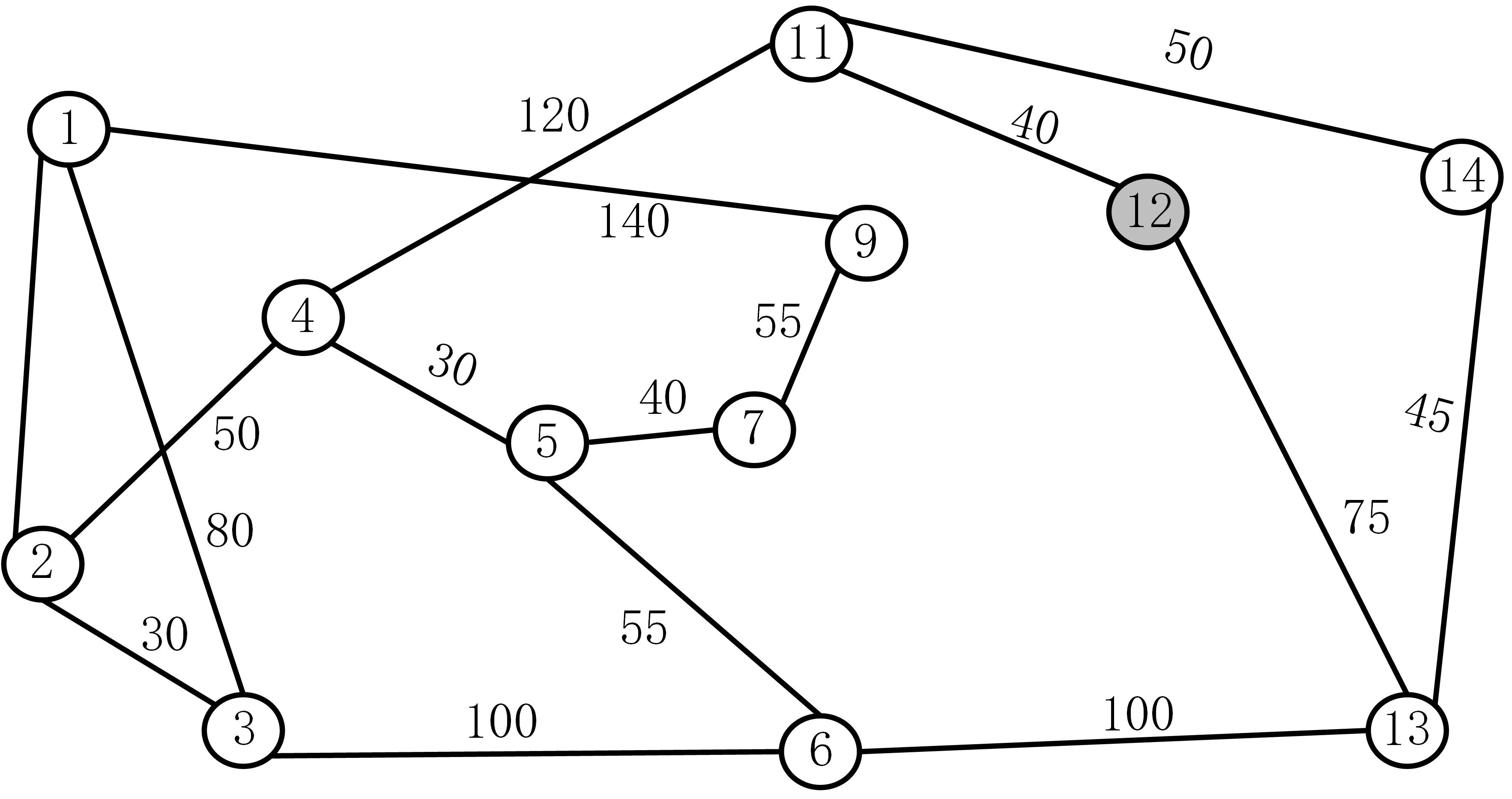}}

	\subfigure[Adding node ${v_8}$ and node ${v_{10}}$]{
	\label{fig:nsfnet5}
	\includegraphics[scale=0.4]{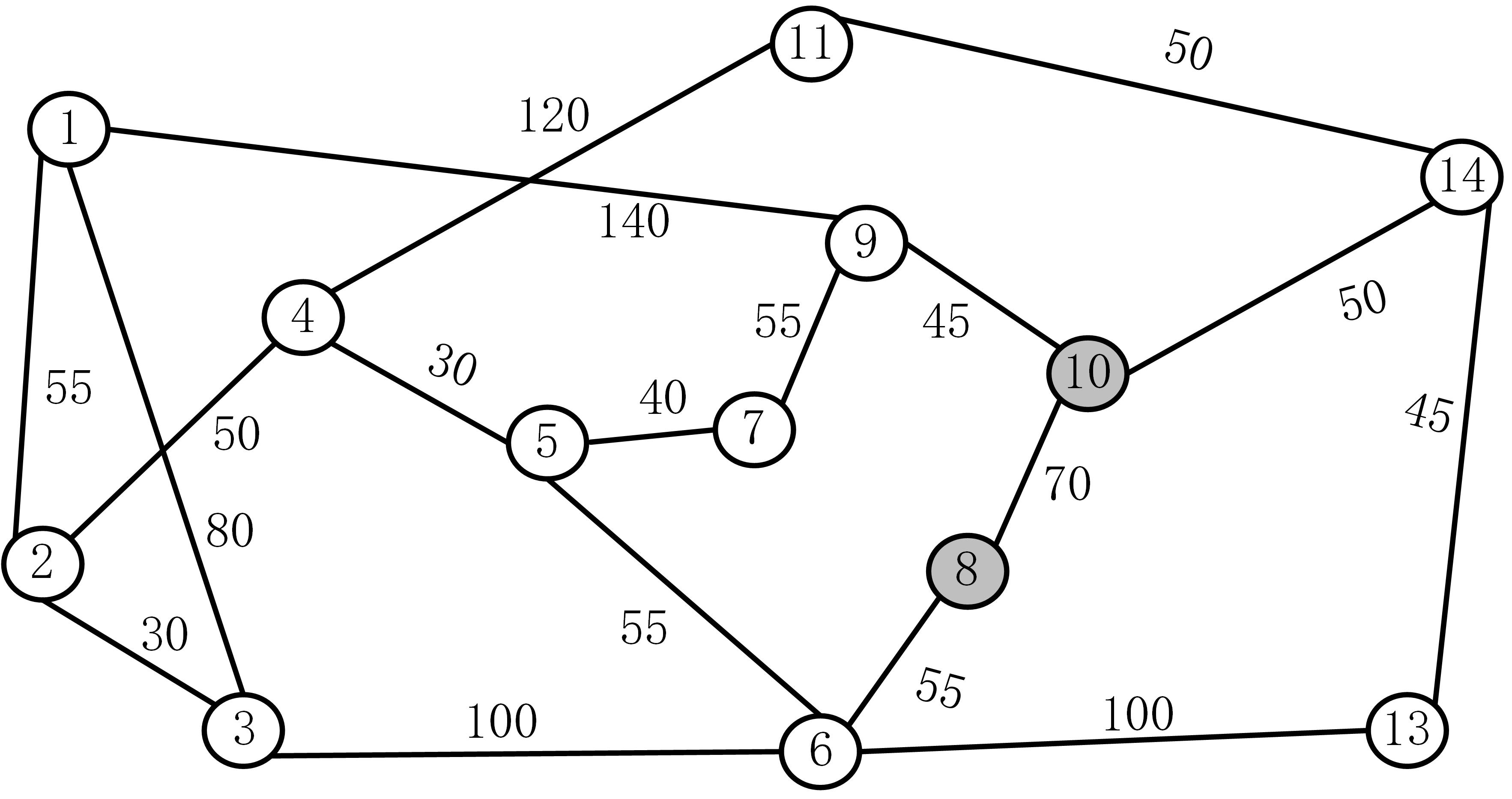}}
	\subfigure[Adding node ${v_8}$ and node ${v_{12}}$]{
	\label{fig:nsfnet6}
	\includegraphics[scale=0.4]{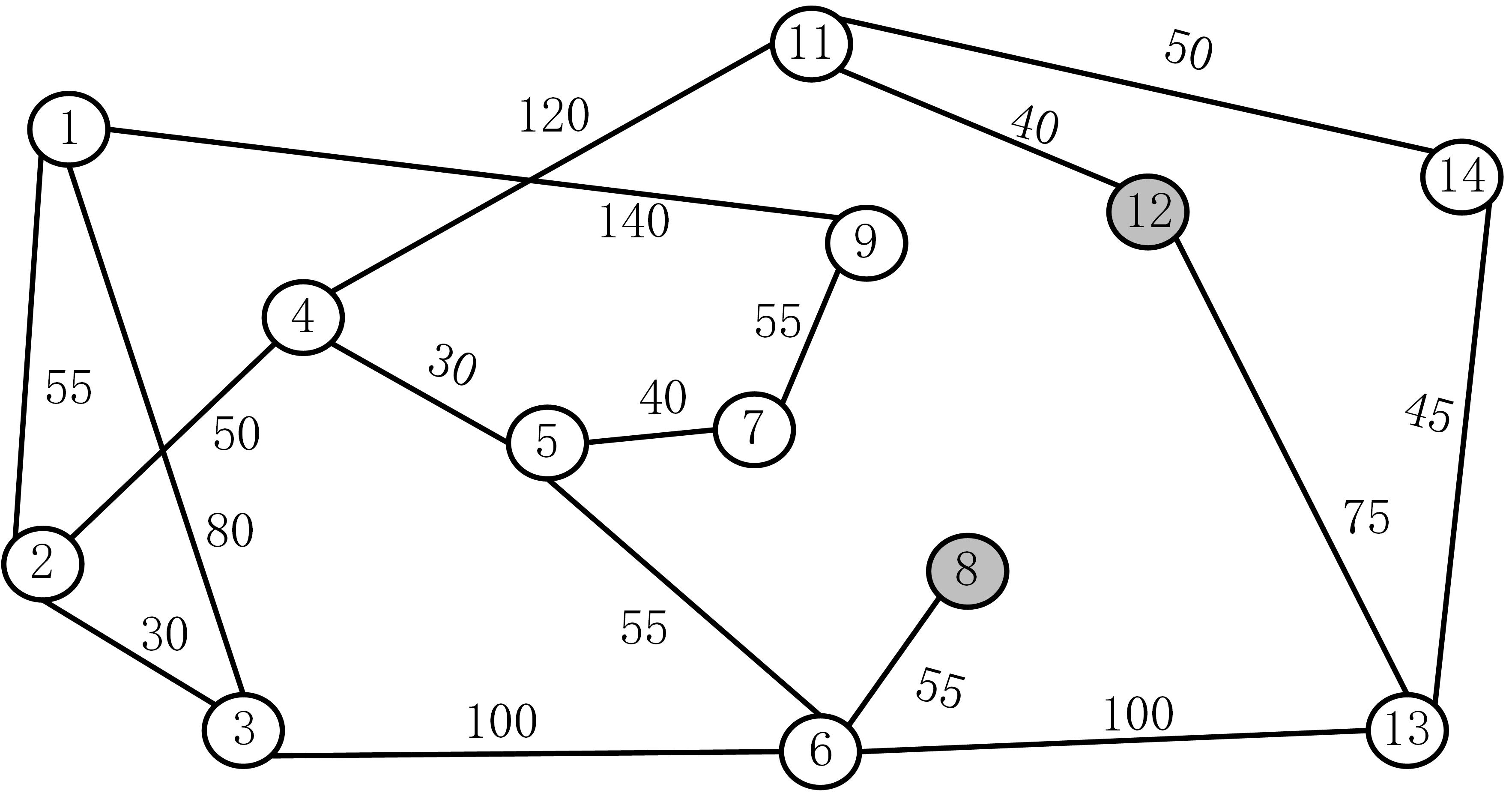}}
	
	\subfigure[Adding node ${v_{10}}$ and node ${v_{12}}$]{
	\label{fig:nsfnet7}
	\includegraphics[scale=0.4]{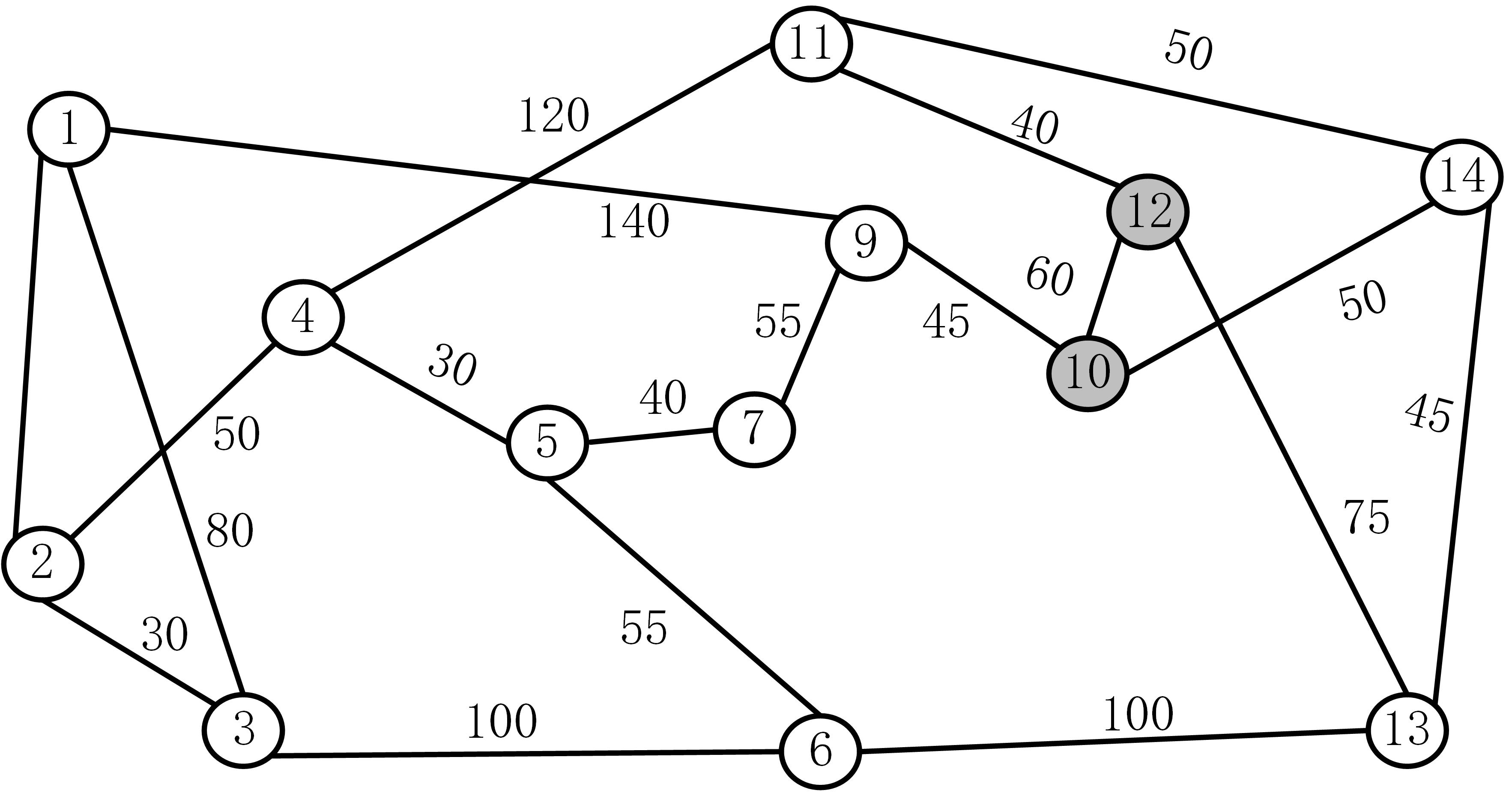}}
	\subfigure[Adding node ${v_8}$,node ${v_{10}}$ and node ${v_{12}}$]{
	\label{fig:nsfnet8}
	\includegraphics[scale=0.4]{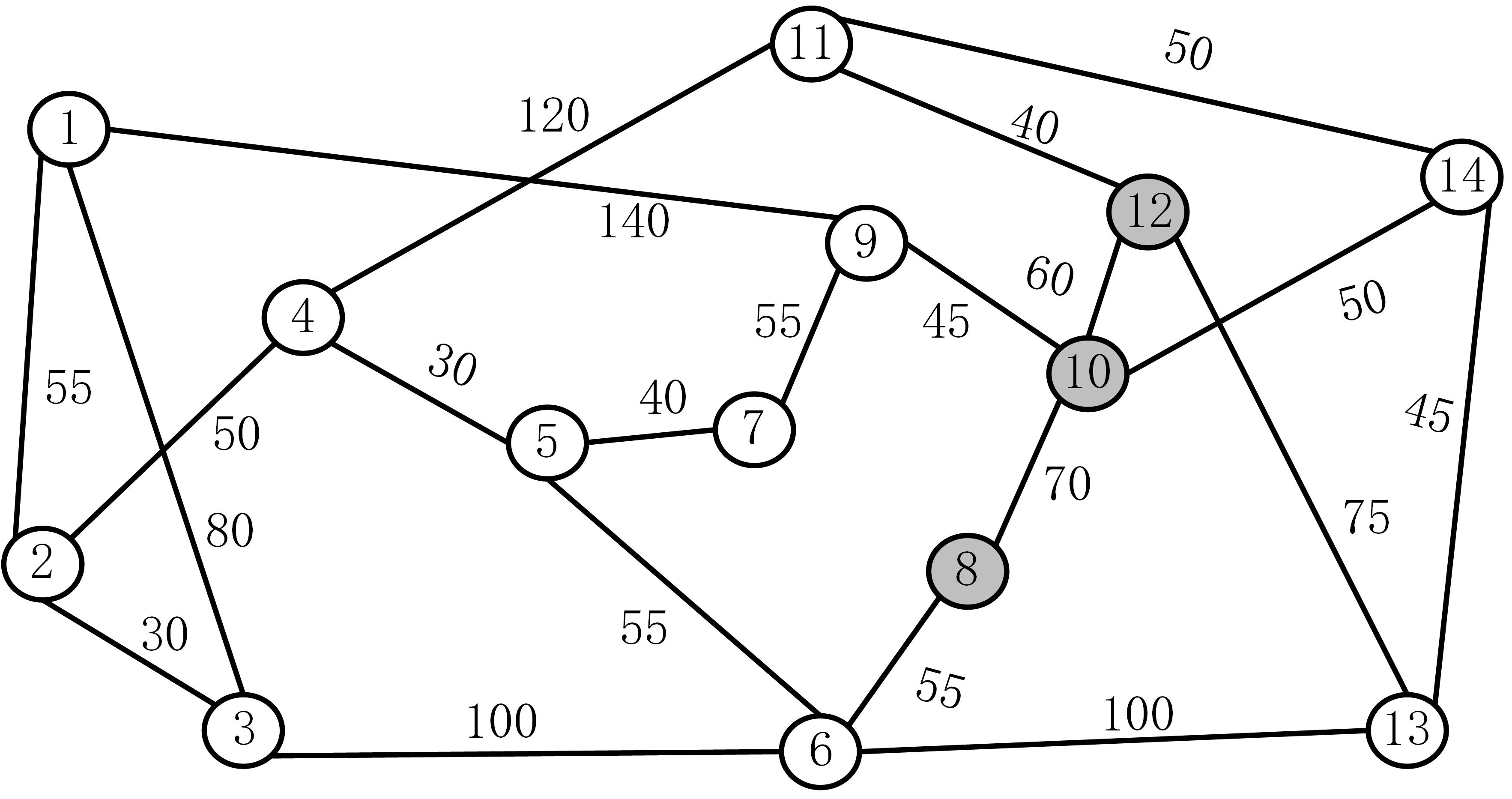}}
\caption{Eight different topologies}
\label{fig_4}
\end{figure}

To determine the optimal selection scheme, different combinations of the three optional nodes form eight different topologies, as shown in Fig. \ref{fig_4}. To quantitatively evaluate different selection schemes, the \textsl{ITS communication bounds} under eight different topologies were calculated. The results are listed Table \ref{tab5}.

\begin{table}[htbp]
	\centering
	\caption{Performance comparison under eight different topologies $\left( {d = 15{\ \rm{Kbps}}} \right)$}
	\label{tab5}       
	\begin{threeparttable}
    \renewcommand\arraystretch{1.5}
    \setlength{\tabcolsep}{1.5mm}{
        \begin{tabular}{ccccccccc}
    		\toprule[1pt]
    		Selection & $none$ & ${v_8}$ & ${v_{10}}$ & ${v_{12}}$ & ${v_8}, {v_{10}}$ & ${v_{8}}, {v_{12}}$ & ${v_{10}}, {v_{12}}$  & $all$ \\
    		\midrule
    		Bound & 0.104 & 0.104 & 1.642 & 0.104 & 2.352 & 0.104 & 1.642 & 2.352 \\
    		\bottomrule[1pt]
    	\end{tabular}
    }
    \end{threeparttable}
\end{table}

For ease of presentation, we use Bound$\left(X\right)$ to denote the value of \textsl{ITS communication bound} when the selection is $X$. As listed in Table \ref{tab5}, Bound$\left(v_{10}\right)$ > 1 and Bound$\left(v_8\right)$ = Bound$\left(v_{12}\right)$ = Bound$\left(none\right)$ < 1.Therefore, to meet the communication demand, only adding node $v_{10}$ is the best selection. In contrast, only adding node $v_8$ or only adding node $v_{12}$ does not yield any performance gains. In addition, Bound$\left(v_8, v_{10}\right)$ > Bound$\left(v_{10}\right)$. That is, on the basis of adding node $v_{10}$, adding node $v_8$ will produce more performance gains. However, Bound$\left(v_{12}, v_{10}\right)$ = Bound $\left(v_{10}\right)$, Bound$\left(v_{12}, v_8\right)$ = Bound $\left(v_8\right)$ and Bound$\left(all\right)$ = Bound$\left(v_8, v_{10}\right)$. Therefore, in any case, the addition of node $v_{12}$ does not bring any performance gains.

In general, only adding node $v_{10}$ can meet the communication demand. On this basis, adding node $v_8$ will further enhance the whole demand satisfaction. However, the addition of node $v_{12}$ does not bring any performance gains. This conclusion cannot be intuitively inferred. Therefore, the necessity of proposed FM model and \textsl{ITS communication bound} indicator is verified.

\section{Conclusion}
\label{sec:conclusion}
This paper has proposed a flow-based mathematical model of a QKD network. The major contributions of this study include: (\uppercase\expandafter{\romannumeral1}) The FM model was proposed by modeling a QKD network as a graph with nodes, edges, and \textsl{QKD-flows}; (\uppercase\expandafter{\romannumeral2}) Based on the created model, a unique QKD network performance indicator was proposed and the corresponding linear programming-based calculation algorithm was designed; (\uppercase\expandafter{\romannumeral3}) The validity and necessity of the proposed FM model and performance indicator were verified through subtly designed simulations addressing two typical topology planning tasks. This study provide us with the means to explore new possibilities in the area of QKD networking and promote the development of QKD networking technology. 

The FM model proposed in this paper can be used for the networking of all kinds of point-to-point QKD protocols, such as BB84-QKD protocol and GG02-QKD protocol. However, some QKD protocols that need to set up a third party, such as MDI-QKD protocol and TF-QKD protocol, cannot be supported. In the future, we will continue to study the networking of such non-point-to-point QKD protocols. In addition, a decoy state discrete-variable QKD protocol was selected in the simulation of this paper. However, it is well known that there are many kinds of QKD protocols that have been put to practical use. In the process of network construction, to control the construction cost, it is necessary to select a reasonable QKD protocol according to the specific application. The FM model can be used to guide the selection of QKD systems by accurately calculating the network topology performance when parameters are set. However, the iterative method for network construction is not sufficiently efficient because it incurs a selection process ultimately leading to the selection of the best case. In our future research, it will be necessary to further consider the relationship between the QKD protocol, link distance, topology, and other factors to propose an efficient and adaptive construction scheme.

\section*{Funding}
Space Science and Technology Advance Research Joint Funds (Grant No.6141B06110105), and
National Natural Science Foundation of China (NSFC) (Grant No. 61771168).

\section*{Disclosures}
The authors declare no conflicts of interest.


\bibliography{reference}






\end{document}